# On the pre-metric foundations of wave mechanics I: massless waves

**D. H. Delphenich**[†]

Lindsborg, KS USA 67456

The mechanics of wave motion in a medium are founded in conservation laws for the physical quantities that the waves carry, combined with the constitutive laws of the medium, and define Lorentzian structures only in degenerate cases of the dispersion laws that follow from the field equations. It is suggested that the transition from wave motion to point motion is best factored into an intermediate step of extended matter motion, which then makes the dimension-codimension duality of waves and trajectories a natural consequence of the bicharacteristic (geodesic) foliation associated with the dispersion law. This process is illustrated in the conventional case of quadratic dispersion laws, as well as quartic ones, which include the Heisenberg-Euler dispersion law. It is suggested that the contributions to geodesic motion from the non-quadratic nature of a dispersion law might represent another source of quantum fluctuations about classical extremals, in addition to the diffraction effects that are left out by the geometrical optics approximation.

## 1 Introduction

In order to define the foundations of wave mechanics, in the more general sense of the mechanics of physical waves in all of their natural manifestations, one must recognize that wave motion comes about as a consequence of deeper assumptions about the nature of the medium in which the waves in question propagate. In particular, whether one is dealing with waves in mechanical media or electromagnetic media, one generally starts with a set of first-order partial differential equations for a field in the medium, which are based in conservation laws for the physical quantities that are associated with the field, and a constitutive law that describes the interaction between the medium and the field. Usually, one can combine the system of first-order partial differential equations and the constitutive law into a single second-order partial differential equation for the field, which will be linear or nonlinear depending upon the nature of the constitutive law.

---

[†] E-mail: david_delphenich@yahoo.com

One then settles upon a specific form for the field that would represent a "wavelike" solution and then derives a dispersion law for that type of wave from the field equation. Various forms of such wave solutions that one regularly considers are time-harmonic field solutions, such as some standing waves, the geometrical optics approximation, plane waves, and waves with other geometric symmetries, such as spherical, ellipsoidal, and cylindrical waves.

Although the mathematical theory of waves usually introduces a Lorentzian structure from the outset, nonetheless, more generally, one finds that the possibility for wave-like fields must exist at this "pre-metric" level of fields and constitutive laws, and in order to account for the appearance of the Lorentzian structure, one must first examine the symbol of the second-order differential operator that defines the field equation. One obtains a linear algebraic operator on wave fields whose determinant defines a characteristic polynomial $F[k]$ in the frequency-wave number covector $k$, which is also called just the *wave covector field*. The vanishing or not of $F[k]$ defines two classes of dispersion laws, which we then call *homogeneous* and *inhomogeneous*, respectively.

One then finds that whether or not the motion of waves in the medium is governed by a Lorentzian structure or something more algebraically involved than a quadratic dispersion polynomial depends entirely upon the nature of the dispersion law, which, in turn, depends largely upon the nature of the constitutive law. A Lorentzian structure usually follows from a degenerate case of a quadratic dispersion law that is associated with a linear homogeneous isotropic medium, while in the more general cases the characteristic polynomial can be quartic or sextic. Hence, one must realize that any medium that supports wave motion has a deeper structure to address before one introduces a Lorentzian structure.

If one treats the relation $k = d\phi$ that expresses the wave covector field $k$ as the differential of a phase function $\phi$, not as a defining identity, but a first-order partial differential equation for $\phi$, then when this is combined with the dispersion law $F[k]$ = const. one finds that one has used the geometrical optics approximation to replace the second-order field equation for the wave function $Ae^{-i\phi}$ with a nonlinear first-order partial differential equation $F[d\phi]$ = const. for the phase function, which gives the shape of the wave fronts; in this approximation, the amplitude function $A$ plays no role, as it is assumed to vary slowly when compared to the time variation of $\phi$.

As long as $F[k]$ is independent of the phase itself, one finds that one is dealing with symplectic geometry of the spacetime cotangent bundle $T^*M$. The characteristic polynomial $F[k]$ plays the role of a Hamiltonian function, although we shall reserve that term for physical case in which one is concerned with energy directly. The characteristic vector field on $T^*M$ that is defined by $F$ and the symplectic structure then generalizes the geodesic flow that would follow from using a Lorentzian metric $g(k, k)$ on $T^*M$ as the source of $F$; in the case of a homogeneous dispersion law, the resulting geodesics would then generalize null geodesics. One finds that there are terms in the geodesic equation that go beyond the Levi-Civita terms and vanish for the case of a quadratic $F$.

Something else that one gets from $F$ and the symplectic structure of $T^*M$ is a generalization of the association of cotangent vectors with tangent vectors that one would obtain from a Lorentzian metric. However, when $F$ is not quadratic, this map from cotangent spaces to their corresponding tangent spaces is not generally a linear isomorphism, as one expects from a metric, but a more involved homogeneous algebraic diffeomorphism, such as a cubic polynomial map. Hence, the relationship between the characteristic hypersurfaces in $T^*M$ and the ones in $T(M)$ is no longer as simple as having both of them be light cones in form or another. However, as we shall see, as long as $F$ is a homogeneous function of $k$ the two hypersurfaces will still be related by a simple rule that has been long known in the quadratic context by the projective geometers. We shall call this relationship between covectors and vectors "dimension-codimension duality," and regard it as describing the transition from wave mechanics to the mechanics of extended matter; i.e., continuum mechanics. That is, the wave covector field $k$ defines isophase hypersurfaces of codimension one that describe the motion of wave fronts, while the associated velocity vector field $\mathbf{v}(k)$ describes the congruence of curves defined by the motion of the individual points of the wave front, and generalizes the geodesics of a Lorentzian structure.

Ordinarily, when one desires to relate the motion of matter waves, as it is described in quantum mechanics, to the motion of points, as it is described in classical mechanics, one generally accomplishes this in one step by virtue of the fact that the statistical interpretation of wave mechanics is based in the assumption that the modulus-squared of the quantum wave function represents a probability density function for the position of a *pointlike* particle. By Ehrenfest's theorem, one then uses this probability density function to define mean values of quantum observables, which are then presumed to describe the classical motion.

Another way of effecting the classical limit of quantum mechanics is to assume that wave mechanics has an analogous relationship to point mechanics as wave optics does to geometrical optics. This suggests that the classical limit is a short-wavelength (high-frequency) limit of wave mechanics, which might correspond to the limit in which Planck's constant $h$ goes to zero.

If one wishes to be more objective about the nature of the density function in question, one might consider that it could also describe a mass density or a charge density. The former case is referred to as the "hydrodynamical" interpretation, while the latter case was suggested by the Pauli-Weisskopf treatment of scalar mesons by the Klein-Gordon equation. Hence, there seems to be a need for an intermediate step between wave mechanics and point mechanics that first converts the motion of waves to the motion of *extended* matter.

One finds that when one is given a dispersion law this transition from wave mechanics to continuum mechanics comes about naturally by way of the bicharacteristic (i.e., geodesic) foliation that is associated with dispersion law. The ultimate passage from continuum mechanics to point mechanics can then be achieved by the method of moments that is commonly used in classical mechanics.

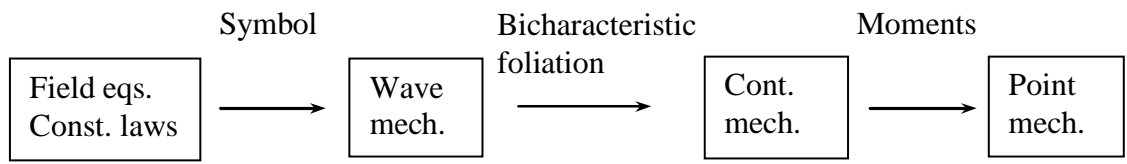

Figure 1. Transition from field equations to point mechanics

We can summarize this sequence of transitions that take us from the basic field equations and constitutive laws to the mechanics of points diagrammatically, as in Figure 1:

So far, we have only been talking about kinematics. It is in the association of energy and momentum with frequency and wave number that one sees that the introduction of the Planck constant is more appropriate to point mechanics than to continuum mechanics, and that more generally one might be dealing with a density that only gives the measured constant *h* when it is integrated over the extended matter distribution.

As long as one is dealing with massless waves, the interpretation of the bicharacteristic foliation as describing the motion of extended material objects is somewhat moot, since one does not have a non-vanishing rest mass density function, or equivalently, a non-vanishing rest frequency. Hence, the method of moments obviously cannot be employed. In a sequel to this article, we shall return to the same flow of ideas that is described here when one wishes to include waves that describe massive matter, which are characterized by inhomogeneous dispersion laws. We shall then see that the appropriate geometry is contact geometry, which is somewhat more general than symplectic geometry, although quite similar.

We now briefly summarize the contents of the remainder of this work. Section 2 is devoted to laying the field-theoretic foundations for wave motion and deriving the resulting dispersion law. In section 3, we discuss the general nature of the geometry that is most intrinsically related to dispersion polynomials in the case of massless waves, namely, symplectic geometry. We also discuss the geometric nature of dimension-codimension duality when the dispersion law is not quadratic. In section 4, we then examine the special forms that the null geodesic equations take for various dispersion laws of interest to electromagnetism.

## 2 From field equations to dispersion laws

In order for wave motion to exist in a physical medium that medium must have a specialized structure that amounts to the association of some sort of "elementary oscillator" to each point of the medium, along with some mechanism that couples each oscillator to the other oscillators, at least in a sufficiently small neighborhood of each point.

Of course, the very definition of an oscillator in its full generality could already lead to excessive complications, since even one-dimensional oscillators can include a broad spectrum of possibilities, such as nonlinearity, damping, and even the absence of strict periodicity, in the sense of a period for the oscillation that does not change in time. For the present purposes, we shall not be concerned with the deeper analysis of the elementary oscillators, though. The mechanism of coupling will take the form of a differential conservation law for a dynamical quantity that is associated with a kinematical quantity by means of a constitutive law.

### 2.1 Field equations

More specifically, one starts with a covector field $\psi$ that is defined on a subset $\mathcal{T} = \mathbf{R} \times \Sigma$ of the spacetime manifold $M$ – whose dimension *n* is at least two – that one calls the *world-tube* of the motion. Note that we shall not introduce the usual causality conditions on the world tube of making it timelike or lightlike since we shall take the position that causality is a *consequence* of wave motion, not a prerequisite for it. That is, the characteristic submanifolds of the tangent spaces to *M* that define the local causal ordering of the points of *M* come about as a corollary to more fundamental considerations about the nature of the medium.

The differential of $\psi$ [1]:

$$D\psi = \psi_{\mu,\nu} \, dx^\mu \otimes dx^\nu \tag{2.1}$$

is a kinematical quantity that plays the role of a generalized velocity. Thus, $D\psi$ will also be defined on the world tube $\mathcal{T}$. For the time being, we shall not pursue the unavoidable introduction of a linear connection in order to make the definition of $D\psi$ independent of the choice of coordinate system, or at least the choice of local coframe. Rather, we shall assume that *M* is an open subset of $\mathbf{R}^n$, so, for now, parallel translation is globally defined by vector translation. Eventually, we will see that parallel translation emerges from the connection that is associated with the dispersion law.

Since $D\psi$ is a second rank covariant tensor field, one can polarize it into a symmetric part $D_s\psi$ and an anti-symmetric part $d\psi$:

$$D_s\psi = \tfrac{1}{2}(\psi_{\mu,\nu} + \psi_{\nu,\mu}) \, dx^\mu \, dx^\nu, \qquad d\psi = \tfrac{1}{2}(\psi_{\mu,\nu} - \psi_{\nu,\mu}) \, dx^\mu \wedge dx^\nu. \tag{2.2}$$

The anti-symmetric part, which is the exterior derivative of the 1-form $\psi$, is a 2-form, and is, moreover, globally defined even in the absence of a linear connection. However, in order to make the symmetric part globally well-defined, one must introduce a linear connection and replace *D* with the covariant differential operator that the connection defines. Since this will not affect the symbol, we shall not insist upon that refinement, at the moment.

Along with $\psi$, we shall define our second fundamental kinematical quantity $\Phi$ to be either $D_s\psi$ or $d\psi$, and refer to a theory based on the former definition to be *mechanical*, while one that is based on the latter is said to be *electromagnetic*. However, the latter case can also describe vorticity waves in a mechanical medium.

In the usual theory of mechanical waves $\psi$ is the displacement covector field $u(x) = (x^i(y) - x^i(x))\partial_i$ that is associated with a deformation $f: U \to M$, $x \mapsto y = f(x)$. The field $\Phi = D_s\psi$ is its corresponding infinitesimal strain tensor field $e = \tfrac{1}{2}(u_{\mu,\nu} + u_{\nu,\mu}) \, dx^\mu \, dx^\nu$.

In Maxwellian electromagnetism, $\psi$ is the electromagnetic potential 1-form $A = A_\mu \, dx^\mu$ so $\Phi = d\psi$ represents the Minkowski field strength 2-form $F = \tfrac{1}{2}(A_{\mu,\nu} - A_{\nu,\mu}) \, dx^\mu \wedge dx^\nu$.

In relativistic hydrodynamics, if $\psi$ is the covelocity 1-form $u = u_\mu \, dx^\mu$ for a fluid motion then $d\psi$ is its *kinematical vorticity* [2] 2-form $\Omega = \tfrac{1}{2}(u_{\mu,\nu} - u_{\nu,\mu}) \, dx^\mu \wedge dx^\nu$, which is also known as its *infinitesimal rate of rotation*.

We then associate either a second rank symmetric contravariant tensor field or a bivector field **B** with $\Phi$ by means of a *constitutive law*. In the mechanical case, this is a diffeomorphism $C_x: S_x^{0,2}(M) \to S_x^{2,0}(M)$ at each $x \in M$ from the vector space $S_x^{0,2}(M)$ of symmetric doubly covariant tensors at $x$ to the vector space $S_x^{2,0}(M)$ of symmetric doubly contravariant tensors at $x$ and the tensor $\mathbf{B}_x = C_x(e_x)$ is the infinitesimal stress tensor $\sigma_x$ at $x$. In the electromagnetic case, it is a diffeomorphism $C_x: \Lambda_x^2(M) \to \Lambda_{2,x}(M)$ from the vector space $\Lambda_x^2(M)$ of algebraic 2-forms at $x$ to the vector space $\Lambda_{2,x}(M)$ of bivectors at $x$, and the bivector $\mathbf{B}_x = C_x(F_x)$ is the electromagnetic excitation bivector $\mathfrak{h}_x$ at $x$.

---

[1] We shall use the notation *Df* for the differential of a map $f: M \to N$, in order to avoid confusion with the exterior derivative operator *d*. Since we will not explicitly introduce the Spencer operator, which also gets customarily notated by *D*, this should not be a source of notational confusion within the present discussion.

[2] We are using the terminology of Carter [**1**] for relativistic hydrodynamics.

If the constitutive law is linear then the diffeomorphism $C$ is assumed to be a linear map in either case. Hence, one will have, in general, the local expressions:

$$B^{\kappa\lambda} = \tfrac{1}{2} C^{\kappa\lambda\mu\nu} \Phi_{\mu\nu}. \tag{2.3}$$

In either the mechanical or electromagnetic case, we can propose that our second fundamental differential equation, besides the definition of $\Phi$, is the conservation law:

$$\delta \mathbf{B} = B^{\mu\nu}_{,\mu} \frac{\partial}{\partial x^\nu} = \mathbf{J}, \tag{2.4}$$

which makes sense regardless of whether $B^{\mu\nu}$ is symmetric or anti-symmetric. The vector field $\mathbf{J}$ on $\mathcal{T}$ represents a source current for the field $\mathbf{B}$, which we denote by $\mathbf{f}$ in the mechanical case to indicate that it then represents a momentum flux or force density.

In summary, we can write our fundamental equations as:

$$e = D_s u, \qquad \delta\boldsymbol{\sigma} = \mathbf{f}, \qquad \boldsymbol{\sigma} = C(e), \tag{2.5}$$

in the mechanical case, and:

$$F = dA, \qquad \delta\mathfrak{h} = \mathbf{J}, \qquad \mathfrak{h} = C(F), \tag{2.5'}$$

in the electromagnetic case. In the mechanical case, $\boldsymbol{\sigma} = \sigma^{ij} \partial_i \partial_j$ is the (symmetric) stress tensor.

In either case, we have a system of first order partial differential equations for $\psi$ and $\mathbf{B}$, combined with a system of algebraic equations that couples $\Phi$ to $\mathbf{B}$. They can be combined into a single second order partial differential equation for $u$ or $A$, resp., which takes the form:

$$\mathbf{f} = \Box_s u \equiv (\delta \cdot C \cdot D_s) u, \tag{2.6}$$

or:

$$\mathbf{J} = \Box_a A \equiv (\delta \cdot C \cdot d) A, \tag{2.6'}$$

respectively. We shall then regard the second order differential operators $\Box_s$ and $\Box_a$ as generalized field operators and refer to them generically by the symbol $\Box_C$.

In local components, if $\psi$ is a 1-form with components $\psi_\mu$ and the constitutive law $C$ takes the most general form with components $C^{\kappa\lambda\mu\nu}(x, \psi)$ then the general field equation takes the form:

$$C^{\kappa\lambda\mu\nu} \frac{\partial^2 \psi_\nu}{\partial x^\lambda \partial x^\mu} + \left( \frac{\partial C^{\kappa\lambda\mu\nu}}{\partial x^\lambda} + \frac{\partial C^{\kappa\lambda\mu\nu}}{\partial \psi_\alpha} \frac{\partial \psi_\alpha}{\partial x^\lambda} \right) \frac{\partial \psi_\nu}{\partial x^\mu} = J^\kappa. \tag{2.7}$$

Note that only the components of $C$ affect the highest-order derivatives in the differential operator $\square_C$, even when $C$ is inhomogeneous or nonlinear.

### 2.2 The representation of wave solutions

Solutions to the systems of partial differential equations in any of the forms that were given above do not have to represent waves. Indeed, the systems may very well reduce to time-invariant systems whose solutions represent equilibrium states or static field configurations. Hence, in order to pass from systems of field equations to wave mechanics, one must first settle upon the exact nature of a wavelike solution to one's field equations.

For the sake of simplicity, we shall assume that the vector space $V$ in which the wave function $\psi: M \to V$ takes its values is a complex vector space. More generally, one might wish to make $\psi$ a section of a complex vector bundle $E \to M$, but unless one is going to address the topological issues that originate in the assumption that this bundle is not trivial, it is entirely sufficient for most physical purposes to consider the trivial case, in which sections are merely vector-valued functions. However, the local issue of invariance under a change of local frame field for the fibers of $E$ is often unavoidable, along with the introduction of a linear connection, if one is to define covariant derivatives of sections.

A particularly important case of wave solutions is given by the *time-harmonic* solutions, which have the form:

$$\psi(t, x^i) = e^{-i\omega t} A(x), \tag{2.8}$$

in which $\omega$ is a real constant that represents the angular frequency of the basic oscillation and $A: \Sigma \to V$ is a function of the spatial position $x \in \Sigma$. Of course, this very construction implies that the spacetime manifold $M$ on which the wave function $\psi$ is defined is *space-time separable*; i.e., $M = \mathbf{R} \times \Sigma$ for some "spatial" manifold $\Sigma$. Of course, making that very construction physically precise and meaningful is fraught with subtleties, but for now we use the concept of space naively.

An important class of examples of time-harmonic waves is given by the class of (time-harmonic) *standing waves*. In such a case, $A(x)$ defines the spatial shape of the wave envelope in which the wave oscillates.

In particular, when one is concerned with linear field equations with constant coefficients that involve fields defined upon the vector space $M = \mathbf{R} \times \mathbf{R}^n$, a particularly popular form for a wave function is given by the *plane wave* solutions:

$$\psi(t, x^i) = A_0 e^{-i\phi(t, x)}, \tag{2.9}$$

in which $A_0 \in V$ is a constant function on $M$. As for the expression $\phi(t, x)$, it takes the local form:

$$\phi(t, x) = k_\mu x^\mu = \omega t - k_i x^i \qquad (\mu = 0, \ldots, n, \; i = 1, \ldots, n), \tag{2.10}$$

in which $\omega$ and $k_i$ are real constants that represent the *angular frequency* of the basic oscillators that support the wave motion and the spatial *wave numbers* of the wave itself, respectively.

One can then define the *frequency-wave number covector* – or simply *wave covector*, for short – by way of:

$$k = k_\mu \, dx^\mu. \tag{2.11}$$

The physical nature of the real function $\phi(t, x)$, which we call the *phase function* for the wave $\psi$, is that its level hypersurfaces, which are then affine hyperplanes in $\mathbf{R} \times \mathbf{R}^n$, represent constant-phase hypersurfaces, which we then call *isophases*. The planes in $\mathbf{R}^n$ that one gets by fixing a value of the time coordinate $t$ are then the *momentary wave fronts*.

Plane-wave solutions are particularly useful in the case of linear field equations with constant coefficients, since one can use the Fourier transform as a means of expressing any other wave function in terms of a linear combination of plane waves. One can even generalize this sort of construction to waves in homogeneous spaces that are more general than affine spaces, such as spheres; surface waves on planets or stars would fall into this category.

It is interesting that even though plane waves are regarded as traveling waves, nevertheless, they are time-harmonic, just as standing plane waves would be. Hence, the distinction between standing waves and traveling waves is not as unambiguous for plane waves as it is for some other types of waves.

A more general form for a wave function is:

$$\psi(t, x) = A(x) \, e^{-i\phi(t, x)}, \tag{2.12}$$

in which the function $A: \Sigma \to V$ is called the *amplitude function*, while the phase function is no longer required to be a linear function of $(t, x)$. Indeed, $\Sigma$ does not have to be a vector space in order for this definition to be meaningful. However, if we assume that $\phi$ approximated by a Taylor series about $(0, 0, \ldots, 0)$ in the form:

$$\phi(t, x^i) = \phi_0 + k_\mu \, x^\mu + \mathcal{O}^2(t, x^i), \tag{2.13}$$

in which:

$$k_\mu = \left.\frac{\partial \phi}{\partial x^\mu}\right|_0, \tag{2.14}$$

then we see that the initial choice of phase $\phi_0$ can be absorbed into the definition of the amplitude function $A(x)$ while the linear term reproduces the tangent plane to the isophase through the origin.

Often the phase function itself is *space-time separable*, in the sense that $\phi(t, x) = \omega t - \phi_s(x)$ for some spatial function $\phi_s(x)$, which then allows us to define a special type of time-harmonic traveling waves (2.8) for which:

$$A(x) = A_0 e^{i\phi_s(x)}. \tag{2.15}$$

The level surfaces of the function $\phi_s(x)$ then give the shape of the traveling wave, such as the aforementioned planes, as well as cylinders, spheres, ellipsoids, and the like. Such a wave is, of course, assumed to have a shape that does not change in time, as opposed to *dispersive* waves, whose shape can change, as well. One sees that the constant amplitude $A_0$ plays no important role in the case of space-time separable waves of the form (2.15), only the phase function.

When one wishes to get away from the necessity of defining waves on linear spaces, or at least, homogeneous spaces, a particularly useful way of representing waves is by means of propagating discontinuities in kinematical or dynamical variables. For instance, shock waves involve finite jump discontinuities in velocity vector fields that are also associated with impulsive jumps in the pressure function when one crosses the *singular hypersurface* over which the jump occurs. This sort of construction goes back to the work of Riemann, Rankine, and Hugoniot on the propagation of waves in gases, and was brought to a mathematical culmination by Hadamard [**2**], who then showed how the construction led into the subtleties of the *Cauchy problem* for the system of partial differential equations that one is concerned with.

The particular form of the Cauchy problem that Hadamard was concerned with involved second-order partial differential equations, so the Cauchy problem was defined by starting with the values of the wave function $\psi$ and its first time derivative $\psi_t = \partial \psi / \partial t$ on an initial (i.e., $t = 0$) hypersurface $S$ in $\mathbf{R} \times \Sigma$. The problem is then to find a unique wave function $\psi$ on $\mathbf{R} \times \Sigma$ that satisfies both the second-order equation and the initial conditions.

Note that as long as the initial wave function $\psi_0(x)$ on $S$ is assumed to be continuously differentiable, one cannot specify the first partial derivatives of $\psi$ with respect to the spatial coordinates on $S$ arbitrarily, which is why one can only specify the initial time derivative. The issue at hand is the extent to which one can specify the second derivatives on $S$, as well, since assuming that $\psi_0(x)$ is $C^2$ in the spatial coordinates will again define those spatial second derivatives uniquely on $S$.

If one is to define an acceleration wave to be a jump discontinuity in the second partial derivative of $\psi$ with respect to $t$ then it is clear that this particular derivative cannot be defined uniquely. When one includes the constraint that $\psi$ must satisfy the field equations on $S$, one finds that $S$ can only be a *characteristic hypersurface* for the field equations, in a sense that we shall clarify in what follows.

### 2.3 Dispersion laws

In order to derive a dispersion law from the field equation, in either form (2.6) or (2.6′) above, we must pass to the "symbol" of the field operator $\Box_C$. Hence, one must understand that dispersion laws are obtained from differential linearizations of more general nonlinear situations.

In classical optics, this approximation is referred to variously as "the limit of geometrical optics," "the high-frequency (short wavelength) limit," and "the eikonal approximation." It starts by assuming that the wave function $\psi$ takes the form (2.12), which then makes:

$$d\psi = e^{-i\phi} dA + i\, d\phi \otimes \psi. \tag{2.16}$$

The approximation takes the form of assuming that the amplitude $A(x)$ is slowly varying – so one can regard $dA$ as essentially null – and a rapidly-varying phase term $\exp(i\phi(x))$, so the differential $d\psi$ takes the form:

$$d\psi = ik \otimes \psi. \tag{2.17}$$

Note that assuming $A$ to be slowly-varying is equivalent to generalizing from plane waves to waves with waves of the form:

$$\psi(t, x) = A_0\, e^{-i\phi(t,\, x)}, \tag{2.18}$$

in which it is not necessary to assume the space-time separability of $M$ in order to make the construction meaningful. Hence, $A_0$ plays not essential role, only $\phi$.

If $\mathcal{D}\colon E \to F$ is a differential operator from sections of a vector bundle $E \to M$ to sections of a vector bundle $F \to M$ then its *symbol* is a bundle map $\sigma[\mathcal{D}]\colon T^*(M) \otimes E \to F$ that takes any $df \otimes s$ to $\mathcal{D}(fs) - f\mathcal{D}(s)$ if $f \in C^\infty(M)$. Hence, it is a linear algebraic map between the fibers $T_x^* \otimes E_x$ and $F_x$ at each point $x \in M$. When one fixes a covector field $k$ the map $\sigma[\mathcal{D}, k]$ takes sections of $E$ to sections of $F$ linearly.

In the case of the differential $D$, since $D(fs) = df \otimes s + fDs$, the symbol of $D$ is tensor multiplication by a covector field $k$:

$$\sigma[D, k](s) = k \otimes s. \tag{2.19}$$

Notice that, except for the factor of $i$, (2.19) is essentially the expression (2.17) that one obtains from the geometrical optics approximation.

By polarization, we then find that:

$$\sigma[D_s, k](s) = k \odot s = \tfrac{1}{2}(k \otimes s + s \otimes k), \tag{2.20a}$$
$$\sigma[d, k](s) = k \wedge s = \tfrac{1}{2}(k \otimes s - s \otimes k). \tag{2.20b}$$

Hence, the linear algebraic operator on sections of $E$ that corresponds to symmetrized or anti-symmetrized differentiation is symmetrized or anti-symmetrized tensor multiplication by $k$, respectively. By abuse of notation, we shall denote any of the three maps $\sigma[D, k]$, $\sigma[D_s, k]$, $\sigma[d, k]$ by $e_k\colon E \to F$ and let the context of its usage dictate the precise meaning implied.

The symbol of the codifferential operator $\delta$ is:

$$\sigma[\delta, k](\Phi) = k(\Phi) = i_k \Phi, \tag{2.21}$$

i.e., interior multiplication by $k$.

In order to find the symbol of the second-order differential operator $\square_C$ we would need to replace $C$ with its linearization $DC$ if it is nonlinear to begin with, so we shall simply assume that $C$ is linear, for brevity. We then get that the total symbol is the composition of the three linear maps:

$$\sigma[\square_C, k] = i_k \cdot C \cdot e_k, \tag{2.22}$$

in either the mechanical or electromagnetic case.

From (2.22), we find that the components of the linear operator $\sigma[\Box_C, k]$ are:

$$\sigma^{\mu\nu}[\Box_C, k] = -C^{\mu\kappa\lambda\nu} k_\kappa k_\lambda. \tag{2.23}$$

In general, the bundle map $\sigma[\Box_C, k]: T^*(M) \to T(M)$ does not have to be invertible, and invertibility will depend upon the choice of $k$. The definition:

$$F[k] \equiv \det \sigma[\Box_C, k] \tag{2.24}$$

then defines a polynomial in $k$ that one calls the *characteristic polynomial* of the differential operator $\Box_C$. It vanishes iff $\sigma[\Box_C, k]$ is not invertible, and the zero locus of $F(k)$ is called the *characteristic hypersurface* for $\Box_C$. We shall call a $k$ that lies in this hypersurface *characteristic*; in the event that $F(k)$ is non-vanishing, we shall call $k$ *non-characteristic*. The algebraic equation in $k$:

$$F[k] = \begin{cases} 0 & k \text{ characteristic} \\ \text{const.} & k \text{ non-characteristic} \end{cases} \tag{2.25}$$

that is defined in either case is then what we call the *dispersion law* for the wave medium in question and the type of wave solution that has been chosen.

It is important to observe that although the classical mathematical theory of wave motion generally only deals with the characteristic case, which corresponds to waves that carry no mass in physics, nonetheless, the mechanics of massive matter waves seems to necessitate the consideration of inhomogeneous dispersion laws and non-characteristic wave vectors. However, since the treatment of non-characteristic wave covectors requirements some subtle revisions to the basic formalism, we shall return to that study in an article that is supplementary to the present one.

From (2.23), one can see that since $k$ appears twice in $\sigma[\Box_C, k]$ this implies that the polynomial $F[k]$ will have a degree that is equal to $2n$. It is, moreover, homogeneous in $k$ of degree $2n$. However, there are some traditional reductions that get applied to the degree.

First, one generally deals with the case of time-invariant constitutive laws, so $M$ takes the space-time separable form $\mathbf{R} \times \Sigma$. This reduces our map $\sigma[\Box_C, k]$ to a map from $T^*(\Sigma)$ to $T(\Sigma)$, and the polynomial $F[k]$ becomes a polynomial $F[\omega, k_i]$ whose degree in $k_i$ is $2(n-1)$. For instance, when $n = 4$ the polynomial in $k$ is homogeneous and sextic.

Second, in the electromagnetic case, the map $e_k: \Lambda^1(\Sigma) \to \Lambda^2(\Sigma)$, $a \mapsto k \wedge a$ is not invertible for any $k$, since its kernel is one-dimensional, namely, all $a$ that take the form $\lambda k$ for some scalar $\lambda \in \mathbf{R}$. This is related to the fact that electromagnetic waves have no longitudinal modes of vibration, but are confined to the *E-B* plane, which is the spatial part of the spacetime 3-plane that is annihilated by $k$. Hence, the characteristic polynomial reduces to a homogeneous quartic polynomial in $k$ in the electromagnetic case.

A case that is of interest in electromagnetism is the *birefringent* case of a dispersion law that is associated with a certain type of anisotropy that is found in "uniaxial" media (see Landau, et al. [**3**]). Birefringence refers to the fact that when $F$ is quartic, if one fixes the spatial components $k_i$ of $k$ then the remaining polynomial $F[\omega, k_i]$ is quadratic in $\omega^2$, and its roots can be shown to be real. This then implies that for any spatial direction of propagation there will generally be two distinct positive values of $\omega$, and therefore two distinct values of the phase velocity $\omega/\kappa$, where $\kappa$ is the Euclidian norm of the spatial covector whose components are $k_i$. This leads to double refraction of a given light ray.

Birefringence is often associated with a factorization of the quartic polynomial $F(k)$ into a product of quadratic polynomials (which is called *bi-metricity* by Barcello, Liberati, and Visser [**4**]), so the completely symmetric fourth-rank covariant tensor that is associated with $F$ takes the form of symmetrized tensor product of two Lorentzian metrics on $T^*(M)$:

$$F = g \odot \tilde{g}. \tag{2.26}$$

The components of $F$ are then obtained from those of $g$ and $\tilde{g}$ by way of:

$$F^{\kappa\lambda\mu\nu} = \tfrac{1}{6}(g^{\kappa\lambda}\tilde{g}^{\mu\nu} + g^{\mu\lambda}\tilde{g}^{\kappa\nu} + g^{\nu\lambda}\tilde{g}^{\mu\kappa} + g^{\kappa\mu}\tilde{g}^{\lambda\nu} + g^{\kappa\nu}\tilde{g}^{\mu\lambda} + g^{\mu\nu}\tilde{g}^{\kappa\lambda}). \tag{2.27}$$

Although fourth-degree polynomials in more than one real variable do not always have to factorize into products of quadratics, nonetheless, in the case of electromagnetic waves, it is widely known that as long as the Lagrangian for the electromagnetic field $F$ depends only upon the Lorentz invariants $F \wedge F$ and $F \wedge *F$, the characteristic polynomial *will* factorize, even in the nonlinear case (see [**4**, **5**]).

A physical example in which non-factorization is the case is given by dispersion laws for "biaxial" optical media, in which the birefringence – i.e., double refraction – is replaced by *conical refraction* (cf., ibid.).

In the case of an *isotropic* medium, the characteristic polynomial reduces to an appropriate power of a quadratic polynomial law and one considers two possible dispersion laws:

$$g^{\mu\nu}k_\mu k_\nu = \begin{cases} 0 & \text{characteristic waves} \\ \omega_0^2 & \text{non-characteristic waves.} \end{cases} \tag{2.28}$$

Since the resulting quadratic form on $k$ that this defines is assumed to be of normal hyperbolic signature type $(+1, -1, \ldots, -1)$, one sees that this is where the light cones[3] finally originate, as well as the unit proper time hyperboloids and mass shells.

Under the traditional Einstein-de Broglie hypothesis that the rest frequency $\omega_0$ is associated with a rest mass $m_0 = \hbar\omega_0/c^2$, we see that characteristic waves are massless, while the matter waves of quantum mechanics must be described by non-characteristic wave covectors, which will be focus of the sequel to this article.

---

[3] Of course, in the mechanical case, they are really *sound cones*.

## 3 Geometric nature of massless wave motion

So far, the dispersion laws that we were considering took the form of smooth functions on the cotangent bundle $T^*(M)$ to a manifold $M$. Actually, we will eventually need to generalize slightly in order to handle non-characteristic waves, and instead of $T^*(M)$ we will consider the manifold $J^1(M, \mathbf{R})$ of all 1-jets of differentiable functions on $M$. This will make it clear that the geometry that is most intrinsic to the description of wave motion is *contact geometry*, which deals with higher-dimensional contact elements than the tangent vectors that one first encounters in differential geometry and point mechanics. However, when dealing with characteristic (i.e., massless) waves it is sufficient to restrict one's scope to the geometry of $T^*(M)$, which is *symplectic geometry*.

Before we recall the rudiments of symplectic geometry, we first discuss the nature of phase foliations that are obtained by starting with the non-zero covector field $k$ and integrating the exterior differential system $k = 0$ that it defines.

### 3.1 Phase foliations

Any non-zero 1-form $k$ defines a sub-bundle of $T(\mathcal{T}) \subset T(M)$ of corank one by its annihilating subspaces. That is, the *algebraic* solution of the exterior differential system:

$$k = 0 \tag{3.1}$$

is a *differential system* on the world-tube $\mathcal{T}$ that is defined by a set of tangent hyperplanes in $T(\mathcal{T})$. One can then examine the integrability of this differential system.

By Frobenius's theorem this differential system is *completely integrable* into a foliation of $\mathcal{T}$ by codimension-one integral submanifolds iff:

$$k \wedge dk = 0 \tag{3.2}$$

at all points of $\mathcal{T}$.

The reason that one includes the adjective "completely" in the previous statement is that it is conceivable that there might be integral submanifolds of the given differential system whose dimension is less than the dimension of the tangent hyperplanes. The issue of finding the maximum dimension for an integral submanifold then becomes important.

A sufficient condition for (3.2) to be true is that $k$ be closed:

$$dk = 0. \tag{3.3}$$

One can think of saying that the *vorticity* of the wave motion vanishes.

By the Poincaré lemma, this is locally equivalent to saying that $k$ is exact:

$$k = d\phi \tag{3.4}$$

for some differentiable function $\phi$ on $\mathcal{T}$, which is unique only up to a locally constant function.

By de Rham's theorem, this local condition is also globally true when $\mathcal{T}$ – hence, $\Sigma$ – is simply connected.

When $k$ is exact, the integral submanifolds of the differential system on $\mathcal{T}$ that is defined by (3.1) are the level hypersurfaces of the phase function $\phi$.

More generally the foliation of $\mathcal{T}$ into "leaves" that consist of the integral submanifolds of the exterior differential system (3.1) will be referred to as a *phase foliation,* and the leaves are then called *isophases*, which generalizes our previous definition. In general, the isophase hypersurfaces in $\mathcal{T}$ will be quite distinct in character from the characteristic hypersurface in $T^*(\mathcal{T})$, even when $\mathcal{T}$ is contained in a vector space, except insofar are the two hypersurfaces share a common tangent hyperplane. For instance, when $\phi(x) = k_\mu x^\mu =$ const. the isophases take the form of affine hyperplanes, which are then tangent to the characteristic hypersurfaces. Similarly, one can give isophases the form of concentric cylinders, spheres and ellipsoids, depending upon the symmetry of the coordinate system that one is using, which is usually traceable to the symmetry of the source of the waves.

As the covector $k$ ranges through all the points of the characteristic hypersurface in a given $T_x^*\mathcal{T}$ it will define a differentiable $n-1$-parameter family of hyperplanes in $T_x\mathcal{T}$, whose envelope is then a characteristic hypersurface in $T_x\mathcal{T}$, by definition. We will discuss the relationship between the two characteristic hypersurfaces in more detail later when we consider what we will be calling "dimension-codimension" duality.

The time evolution of the initial submanifold $\Sigma$ in $\mathcal{T}$ defines a foliation of $\mathcal{T}$ into leaves that are each diffeomorphic to $\Sigma$; they are the level surfaces of the submersion $\mathbf{R} \times \Sigma \to \mathbf{R}$, $(t, x) \mapsto t$, and are referred to as *simultaneity hypersurfaces*, or *isochrones*. More generally, one can consider level hypersurfaces of a proper-time function $t: M \to \mathbf{R}$ whose differential $dt$ is non-vanishing, although the nature of solutions to the Cauchy problem generally implies the existence of a product decomposition $\mathbf{R} \times \Sigma$ for $\mathcal{T}$.

The intersections of the leaves of these two foliations are then the *momentary wave fronts*, which also generalizes the previous definition.

One sees that when $\mathcal{T}$ is contained in a vector space, there exists a class of *elementary waves* for a given polynomial $F[k]$, which then amounts to isophase hypersurfaces that are obtained by projecting the characteristic hypersurface in a typical tangent space onto the vector space itself. For instance, in Minkowski space ($\mathbf{R}^4$, $\eta_{\mu\nu}$) a light cone in any tangent space $T_x\mathbf{R}^4$ projects into a light cone in $\mathbf{R}^4$ through the point $x$, and this light cone represents an expanding spherical momentary wave front whose source is the point $x$, *as long as one assumes that the constitutive properties of the medium are spatially homogeneous.* In the inhomogeneous case, the elementary waves have an idealized character that relates only to the local behavior of wave propagation.

We now see that what the transition from the original field operator to its symbol has accomplished is to replace the second-order field equation $\square_C \psi = 0$ with the pair of equations:

$$F[k] = 0, \qquad k = 0, \tag{3.5}$$

the first of which is algebraic, but nonlinear, and the second of which is an exterior differential system whose integral submanifolds define the isophase hypersurfaces. The generalized eikonal equation:

$$F[d\phi] = 0 \tag{3.6}$$

is a nonlinear first-order partial differential equation for $\phi$ that represents the form that (3.5) takes when one adds the integrability condition that $k = d\phi$.

### 3.2 Phase velocity

Although the fundamental object in wave kinematics is the wave covector $k$, instead of a velocity vector field, as one would consider in fluid mechanics, one can still associate a spatial velocity with $k$ by way of the *phase velocity*. However, one must note that there is something subtle going on that is not usually mentioned, namely, that the "space" that one is talking about is not a tangent *vector* space at each point $x \in M$, but a (co)tangent *projective space*, namely, the projectivized cotangent space $PT_x^*M$ that one obtains from $T_x^*M$ by considering the set of all lines through the origin. That is, a point $[\alpha] \in PT_x^*M$ is the set of all non-zero scalar multiples $\lambda\alpha$ of a non-zero covector $\alpha \in T_x^*M$.

If $\{\theta^\mu, \mu = 0, \ldots, n\}$ is a linear coframe for the vector space $T_x^*M$, so the components of a covector $\alpha \neq 0$ with respect to this coframe are $\alpha_\mu$, then one can also refer to the ordered $n+1$-tuple $(\alpha_0, \ldots, \alpha_n)$ as the *homogeneous coordinates* of the corresponding point $[\alpha]$ in $PT_x^*M$. As long as $\alpha_0 \neq 0$, one can then define the ordered $n$-tuple $(A_1, \ldots, A_n)$ with $A_i = \alpha_i / \alpha_0$ to be the *inhomogeneous (or Plücker) coordinates* of $[\alpha]$. Actually, since the $n$-dimensional projective space $PT_x^*M$ is projectively equivalent to $\mathbf{R}P^n$ (the set of all lines through the origin of $\mathbf{R}^{n+1}$), which is compact, as well as non-orientable, one cannot cover it with just one set of coordinates. Indeed, one needs to use the other $n$ inhomogeneous coordinate systems that one obtains by choosing each other non-zero homogeneous coordinate $\alpha_i$ in succession in place of $\alpha_0$.

Now let us apply this construction to the non-zero 1-form $k = k_\mu dx^\mu$. If we regard $(\omega, k_1, \ldots, k_n)$ as the homogeneous coordinates of the point $[k] \in PT_x^*M$ then the corresponding inhomogeneous coordinates $(s_1, \ldots, s_n)$, with:

$$s_i = \frac{k_i}{\omega} \tag{3.7}$$

define what is sometimes called the *slowness covector* [6] associated with $k$. Of course, since $PT_x^*M$ is not a vector space, but a projective space, one cannot define linear combinations, but one might still think of the $s_i$ as describing the *projective* 1-form $[k]$, which represents the *hyperplane* $k = 0$ in $T_xM$. One then considers projective classes of covectors, instead of the covectors themselves, since replacing $k$ with $\lambda k$ for a non-zero scalar $\lambda$ will not affect the isophase foliation, as one sees from (3.5). This depends crucially upon the fact that we have used a homogeneous polynomial for $F[k]$ and looked at its zero locus, instead of a level hypersurface for a non-zero value of $F[k]$.

One now sees that the $s_i$ that we have introduced in a geometrically natural way are really the reciprocals of the components:

$$v_p^i = \frac{\omega}{k_i} \qquad (3.8)$$

of what is usually called the *phase velocity* vector. However, since taking the reciprocals of components of vectors is hardly a geometrically natural operation, we see that it is the slowness covector that seems to play the fundamental role. Although one could think of the coordinates $s_i$ as having the character of indices of refraction, this association would only be true in special cases of dispersion laws, since the principal indices of refraction are associated with the choice of dispersion law by way of Fresnel analysis [**3**]; we shall more to say about this shortly.

Note that, except for the constraint $F[k] = 0$ that we have imposed on the covector field $k$, the definition of phase velocity has nothing to do with the nature of $F$ itself, and can be defined for any non-zero $k$. Later, we will show how the slowness covector relates to the group velocity vector, which is specific to the choice of $F$ and has an analogous projective-geometric definition.

### 3.3 Symplectic geometry

The manifold $T^*M$ has local coordinate charts of the form $(x^\mu, k_\mu)$, so a 1-form $\alpha$ – i.e., a covector field – on $T^*M$ takes the local form:

$$\alpha = \alpha_\mu \, dx^\mu + \alpha^\mu \, dk_\mu. \qquad (3.9)$$

Note that all of the component functions will be functions on $T^*M$, so their independent variables will locally be $(x^\lambda, k_\lambda)$. However, when one chooses a covector field $k: M \to T^*M$, one can always pull the 1-form $\alpha$ down to a 1-form $k^*\alpha$ on $M$ whose local component form is $\alpha_\mu (x^\lambda, k_\lambda(x)) \, dx^\mu$.

The manifold $T^*M$ is equipped with a canonical 1-form $\theta$ that derives from the fact that any covector on $M$ pulls back to a covector on $T^*M$ under the projection $T^*M \to M$. This canonical 1-form is easiest to understand in its local form:

$$\theta = k_\mu \, dx^\mu. \qquad (3.10)$$

Hence, if $k: M \to T^*M$, $x \mapsto k(x)$ is a 1-form on $M$ then the canonical 1-form $\theta$ will pull down to $k_\mu(x) \, dx^\mu$, which is to say, $k$ itself.

A vector field $X$ on $T^*M$ will take the local form:

$$X = X^\mu \frac{\partial}{\partial x^\mu} + X_\mu \frac{\partial}{\partial k^\mu}, \qquad (3.11)$$

in which all of the components are functions on $T^*M$,.

The second term in this decomposition:

$$X_v = X_\mu \frac{\partial}{\partial k_\mu} \qquad (3.12)$$

can be defined independently of the choice of coordinate system, and is called the *vertical* part of $X$. In general, a vertical tangent vector on $T^*M$ projects to zero under the differential of the projection $\pi: T^*M \to M$.

Locally, the components $(X^\mu, X_\mu)$ of a tangent vector $X$ on $T^*M$ that is annihilated by $\theta$ will satisfy:

$$k_\mu X^\mu = 0, \qquad X_\mu \text{ arbitrary}. \tag{3.13}$$

The set of all such hyperplanes defines a corank-one sub-bundle $C(T^*M)$ of $T(T^*M)$ which then amounts to a differential system on $T^*M$. By Frobenius's theorem, its complete integrability into a codimension-one foliation of $T^*M$ by integral submanifolds is governed by the vanishing of the 3-form:

$$\theta \wedge d\theta = - k_\mu \, dk_\nu \wedge dx^\mu \wedge dx^\nu . \tag{3.14}$$

Since this does not apparently vanish identically, we conclude that in general the sub-bundle $C(T^*M)$ is not completely integrable.

One can next define a canonical 2-form $\Omega$ on $T^*M$ by way of:

$$\Omega = d\theta = dk_\mu \wedge dx^\mu . \tag{3.15}$$

Since this makes $d\Omega = 0$, as well as $\Omega \wedge d\Omega$, we see that the exterior differential system $\Omega = 0$ is completely integrable; indeed, by definition, $\Omega$ is exact. The dimension for an integral submanifold of this exterior differential system is given by the *rank* of the 2-form $\Omega$. There are various ways of defining this concept, but the ones that we shall use are the minimum number of 1-forms on $T^*M$ that are necessary to express the 2-form $\Omega$ and the minimum integer $r$ such that $\Omega \wedge \ldots \wedge \Omega = 0$ when the exterior product has $r$ factors.

This rank turns out to be $n$ ($= \dim M$), and such integral submanifolds of maximal dimension are called *Lagrangian submanifolds* [**6**]. A section $k: M \to T^*M$ will define a Lagrangian submanifold iff $k^*\Omega = 0$. This means:

$$0 = k^*\Omega = k^*d\theta = dk^*\theta = dk. \tag{3.16}$$

Hence, the section $k$ defines a Lagrangian submanifold of $T^*M$ iff it is closed.

One sees that $\Omega$ defines a symplectic structure on the tangent hyperplanes in $T(T^*(M))$ since it is a closed and non-degenerate 2-form. Non-degeneracy implies that there is a canonical linear isomorphism of each $T_k(T^*(M))$ with its dual $T_k^*(T^*(M))$ that takes a tangent vector $X$ at $(x, k)$ to the covector $i_X\Omega$, which locally looks like:

$$i_X\Omega = X_\mu \, dx^\mu - X^\mu \, dk_\mu . \tag{3.17}$$

### 3.4 Characteristic vector field of a function on $T^*M$

If one is given a differentiable function $F: T^*M \to \mathbf{R}$ then one can immediately define a 1-form on $T^*M$ by way of $dF$. Since there is a canonical linear isomorphism of each cotangent space in $T^*(T^*M)$ with its corresponding tangent space $T(T^*M)$, we can associate a canonical vector field $X_F$ on $T^*M$ with $dF$ directly by way of:

$$i_{X_F}\Omega = dF. \tag{3.18}$$

Locally, this equation takes the form:

$$X_\mu\, dx^\mu - X^\mu\, dk_\mu = F_{,\mu}\, dx^\mu + F^{,\mu}\, dk_\mu = F_{,\mu}\, dx^\mu + F^{,\mu}\, dk_\mu. \tag{3.19}$$

We then deduce that the components of $X_F$ are obtained from:

$$X^\mu = F^{,\mu}, \qquad X_\mu = -F_{,\mu}. \tag{3.20}$$

We call the vector field $X_F$ that is associated with $F$ in this way the *characteristic vector field* for $F$. If one chooses to regard the function $F$ as a Hamiltonian function for a conservative mechanical system then it is also customary to refer to the characteristic vector field $X_F$ as the *Hamiltonian vector field* associated with $F$. However, since we are still dealing with wave kinematics, not wave dynamics, it would be confusing to use that terminology at this point. Of course, in conventional wave mechanics the difference between the energy-momentum 1-form and the wave covector amounts to the multiplicative universal constant $\hbar$, but we shall have more to say about that issue in the second part of this study.

Since any vector field on a manifold defines a system of first-order ordinary differential equations by way of the velocity vectors to its integral curves, one then has the following system of equations:

$$\frac{dx^\mu}{d\tau} = \frac{\partial F}{\partial k_\mu}, \qquad \frac{dk_\mu}{d\tau} = -\frac{\partial F}{\partial x^\mu}, \tag{3.21}$$

in which we are using $\tau$ for the curve parameter.

Because the functions $F$ that we shall be concerned with are already "characteristic" for the second-order field operator $\square_C$, we shall refer to the one-dimensional foliation of $T^*M$ by the integral curves of $X_F$ as the *bicharacteristic foliation* for $F$. (Since the term "characteristic" is used in the context of both first-order and second-order differential equations, one must be careful to point out when the terminology is and is not consistent.) Equations (3.21) are then referred to as the system of *bicharacteristic* equations, and amount to the canonical – or Hamilton – equations that are associated with the Hamiltonian $H$.

For a given $F$, the local flows of the vector field $X_F$ will have the property that they preserve the 2-form $\Omega$ – hence, the symplectic structure, – and are therefore referred to as *canonical transformations*, so any characteristic vector field is the infinitesimal generator of a one-parameter family of canonical transformations. To verify the asserted property of $X_F$, one need only take the Lie derivative of $\Omega$ with respect to $X_F$, using Cartan's formula for $L_X$:

$$L_X\Omega = i_X d\Omega + d i_X\Omega = d i_X\Omega = d(dF) = 0. \tag{3.22}$$

(Here, we suppress the subscript $F$ for ease of notation.)

The system (3.21) of ordinary differential equations is also defined by the initial-value problem associated with the first-order partial differential equation in the function $\phi$ on $M$:

$$F[d\phi] = F(x^\mu, \phi_{,\mu}) = 0. \tag{3.23}$$

when one uses Cauchy's method of characteristics (see Duff [8], Arnol'd [9], Courant and Hilbert [10]); this equation is one form of the *Hamilton-Jacobi* equation for the function $\phi$, as well as the eikonal equation when $F$ represents a dispersion polynomial.

An initial-value problem for this first-order equation – which is also a form of the Cauchy problem – is defined by fixing $x^0 = \tau_0$ and a function $\phi_0(\tau_0, x^i)$ on the initial submanifold $\Sigma$ in $M$ that is locally defined by $x^0 = \tau_0$. One then looks for a solution $\phi$ to (3.23) that is defined for all other values of $x^0$ and agrees with $y_0$ on the initial submanifold, while being constant on the bicharacteristic curves; i.e., $\phi(\tau, x^i(\tau)) = \phi(\tau_0, x^i(\tau_0)) = \phi_0(\tau_0, x^i)$. Hence, a solution will always be defined on a world tube of the form $\mathbf{R} \times \Sigma$. In effect, the solution of a partial differential equation gets converted into the solution of a system of ordinary differential equations. One also sees that the initial submanifold $\Sigma$ cannot be tangent to the characteristic hypersurface, or else the bicharacteristic flow would take points of the initial submanifold to other such points. Hence, one must define the initial-value problem on a "non-characteristic" submanifold of $M$. The points of the initial submanifold will then propagate along the bicharacteristics in a manner that sweeps out $\mathbf{R} \times \Sigma$. However, one must note that the submanifold $d\phi: M \to T^*M$ *is* tangent to the characteristic vector field $X_F$ on $T^*M$.

### 3.5 Dimension-codimension duality

One of the advantages of defining a Lorentzian metric $g$ on $T(M)$ is that it allows one to define a linear isomorphism from each tangent space $T_xM$ to each corresponding cotangent space $T^*_xM$ that takes a tangent vector $\mathbf{v} \in T_xM$ to a tangent covector $i_{\mathbf{v}}g = (g_{\mu\nu}v^\nu)\,dx^\mu$; often, this isomorphism is referred to as "lowering the index of $v^\mu$."

The question arises of how to generalize this process when one starts with a characteristic polynomial $F[k]$ instead of a Lorentzian metric. As it turns out, one can still define a *diffeomorphism* of each $T_xM$ with $T^*_xM$, but if the degree of $F[k]$ is $r$ then the diffeomorphism will be a homogeneous polynomial homogeneous of degree $r-1$, which will only be linear when $r = 2$; we shall have more to say about homogeneous function the next section.

In order to define this map, consider the sequence of maps:

$$T^*(M) \xrightarrow{X_F} T(T^*) \xrightarrow{D\pi} T(M).$$

The composition of the two maps is then $(D\pi \cdot X_F): T^*(M) \to T(M)$, $k \mapsto D\pi(X_F(k))$.

Hence, if $k$ is a covector in $T^*_xM$ then we are associating a vector $\mathbf{v}(k)$ in $T_xM$ whose local components take the form:

$$v^\mu = \frac{\partial F}{\partial k_\mu}(x,k). \tag{3.24}$$

Similarly, a covector field $k(x)$ on $M$ goes to a vector field $\mathbf{v}(x)$ that has the local components:

$$v^\mu(x) = \frac{\partial F}{\partial k_\mu}(x, k(x)). \tag{3.25}$$

Since 1-forms – i.e., covector fields – on a world tube $\mathcal{T}$ define codimension-one foliations – when they are integrable – and vector fields, or rather, line fields – which are always integrable – define one-dimensional foliations of $\mathcal{T}$, we that this association of a 1-form on $\mathcal{T}$ with a vector field amounts to a sort of "dimension-codimension" duality between the foliations in question.

This association of lines and hyperplanes is distinct from the Poincaré duality #: $T(M) \to \Lambda^{n-1}(M)$, $\mathbf{v} \mapsto i_\mathbf{v}\mathcal{V}$ that one gets from choosing a volume element on an orientable $M$. This duality, which is not dependent upon any choice of $F$, is based in the fact that a given tangent line can either be generated by a single non-zero tangent vector or annihilated by $n-1$ linearly independent tangent covectors, which then collectively define a non-zero $n-1$-form by exterior multiplication.

We can use dimension-codimension duality for a choice of $F$ to define the aforementioned characteristic hypersurface in $T_x\mathcal{T}$ that corresponds to the characteristic hypersurface $F[k] = 0$ in $T_x^*\mathcal{T}$ by defining the function $G[\mathbf{v}]$ on $T(T)$ to be:

$$G[\mathbf{v}(k)] = F[k]. \tag{3.26}$$

The characteristic hypersurface in $T_x\mathcal{T}$ is then defined by $G[\mathbf{v}] = 0$.

In the geometrical optics of electromagnetic waves, the hypersurface $F[k] = 0$ in $T_x^*\mathcal{T}$ - or rather, its image in $PT_x^*\mathcal{T}$ – defines what amounts to a generalization of the *Fresnel wave (or normal) surface*, while the characteristic hypersurface $G[\mathbf{v}] = 0$ in $PT_x\mathcal{T}$ defines the corresponding *Fresnel ray surface.* In the case of an electromagnetically linear, isotropic, and homogeneous medium, the hypersurfaces are both the familiar light cones of relativity theory. In the next section, we shall see that some of their properties still persist when $F$ is homogeneous, but not necessarily quadratic.

Physically, we are suggesting that if the fundamental kinematic objects of wave mechanics are isophases then the passage to the corresponding bicharacteristic curves is another form of the geometrical optics approximation. Hence, one should probably regard the geometry of waves as being more physically fundamental than the geometry of curves, which pertains to geometrical mechanics, in the same way that wave optics is potentially richer in scope than geometrical optics.

### 3.6 Group velocity

When a specific choice of $F$ has been made, dimension-codimension duality associates a velocity vector $\mathbf{v}(k)$ in $T_xM$ with each wave covector $k$ in $T_x^*M$. The components of $\mathbf{v}(k)$ with respect to a natural frame field are $v^\mu(k_\nu) = \partial F / \partial k_\mu$ when $k = k_\mu\, dx^\mu$.

It is interesting to carry out a procedure that is analogous to the one that we described above in section 3.2 that resulted in the slowness covector. That is, now, instead of considering the projective space $PT_x^*M$ of all lines through the origin in $T_x^*M$, we consider the projective space $PT_xM$ of all lines though the origin in $T_xM$. Just as $T_x^*M$ is the dual of the vector space $T_xM$, similarly, the projective space $PT_x^*M$ is dual to the projective space $PT_xM$, which is also projectively equivalent to $\mathbf{R}P^n$. Now, if $\mathbf{v} = v^\mu \partial_\mu$ is a non-zero tangent vector in $T_xM$ then we regard the $n+1$-tuple $(v^0, \ldots, v^n)$ as the homogeneous coordinates of the line $[\mathbf{v}]$ in $PT_xM$ and $(-v_g^1, \ldots, -v_g^n)$ with $v_g^i = -v^i/v^0$ as its inhomogeneous coordinates.

The reason that we include the minus sign is that when one substitutes $\partial F/\partial \omega$ for $v^0$ and $\partial F/\partial k_i$ for $v^i$ one finds that what one is dealing with is:

$$v_g^i = -\frac{\partial F/\partial k_i}{\partial F/\partial \omega} = \frac{\partial \omega}{\partial k_i}, \tag{3.27}$$

as long as one can solve the dispersion law $F[k] = 0$ for $\omega$, which, by the implicit function theorem, is equivalent to the assumption that $v^0$ is non-zero; however, this is also necessary if we are to define the $v_g^i$, to begin with. Now, we see that the inhomogeneous coordinates $v_g^i$ for the tangent line $[\mathbf{v}]$ also define what is commonly called the *group velocity* vector [**11**] for the wave motion, although we now see that what we are dealing with is a line, not a vector. Unlike the phase velocity, which is defined by $k$ alone, the group velocity vector depends upon the choice of dispersion law, and thus upon the particular constitutive law of the medium.

Just as the isophase foliation is defined by the projective equivalence class $[k]$, and not by the specific choice of representative covector $k$, similarly, the null geodesic congruence that is obtained by the integrating the velocity vector field $\mathbf{v}$ defined by the projective equivalence class $[\mathbf{v}]$ and not the specific choice of $\mathbf{v}$, as long as it is non-zero. Hence, the integral submanifolds in either case are sensitive to the projective geometry of hyperplanes and lines, not the linear geometry of covectors and vectors, respectively. In the case of a line field $[\mathbf{v}(x)]$, the effect of changing the representative vector field $\mathbf{v}(x)$ is to change the parameterization of the curve, but not the set of points that it traces out.

One finds that there is an important relationship between $s_i$ and $v_g^i$ that is true for any homogeneous dispersion polynomial $F$. First, let us review some of the elementary facts about homogeneous functions on vector spaces.

By definition, a function $F: V \to \mathbf{R}$, where $V$ is a vector space, is called *homogeneous of degree r* iff for any scalar $\lambda \in \mathbf{R}$, one has:

$$F(\lambda \mathbf{v}) = \lambda^r F(\mathbf{v}) \tag{3.28}$$

for every vector $\mathbf{v} \in V$.

Examples of homogeneous functions of degree $r$ include homogeneous polynomials of degree $r$, the quotient of a homogeneous polynomial of degree $r+k$ over a homogeneous polynomial of degree $k$, and the $1/k^{\text{th}}$ power of a homogeneous polynomial of degree $rk$. In particular, any linear functional on $\mathbf{R}^3$ is homogeneous of degree 1, but the usual Euclidian norm $\sqrt{x^2 + y^2 + z^2}$ represents a homogeneous function that is of degree 1, but not linear.

A fundamental property of any homogeneous function $f$ – of any degree – on a vector space $V$ is that since it will take the same value $f(\mathbf{v})$ all along the line $[\mathbf{v}]$ as it does for the non-zero vector $\mathbf{v}$, $f$ will define a function $f[\mathbf{v}]$ on the projective space $PV$ of all lines through the origin on $V$ by setting $f[\mathbf{v}] = f(\mathbf{v})$ for any $\mathbf{v}$ that generates $[\mathbf{v}]$. Conversely, under the projection $V - \{0\} \to PV$, $\mathbf{v} \mapsto [\mathbf{v}]$, any function on $PV$ can be pulled back to a homogeneous function in $V$. This is most easily expressed in terms of the homogeneous and inhomogeneous coordinates for $PV$:

$$f(x^0, \ldots, x^n) = \tilde{f}(X^1, \cdots, X^n) = \tilde{f}\left(\frac{x^1}{x^0}, \cdots, \frac{x^n}{x^0}\right). \tag{3.29}$$

For example, the inhomogeneous quadratic polynomial $aX^2 + bX + c$ corresponds to the homogeneous one $ax^2 + bxy + cy^2$ when one sets $X = x/y$. The Minkowski space light cone $c^2t^2 - r^2 = 0$ becomes the sphere $R = 1$ when one regards Minkowski space as the space of homogeneous coordinates for $\mathbf{RP}^3$ and sets the inhomogeneous coordinate equal to $X^i = x^i/ct$, which then represents a sphere of radius one light-second if one regards the space of observation as $\mathbf{RP}^3$.

One can show that a homogeneous function must be continuously differentiable and that when one differentiates (3.28) with respect to $\lambda$, one can deduce *Euler's formula:*

$$F(\mathbf{v}) = \frac{1}{r} dF(\mathbf{v}), \tag{3.30}$$

or, in $k_\mu$ coordinates on $V$:

$$F(k_\mu) = \frac{1}{r} \frac{\partial F}{\partial k_\mu} k_\mu. \tag{3.31}$$

Since this implies that $dF$ is homogeneous of degree $r - 1$ – hence, differentiable – a further iteration of Euler's formula gives:

$$\frac{\partial F}{\partial k_\mu} = \frac{1}{r-1} \frac{\partial^2 F}{\partial k_\mu \partial k_\nu} k_\nu \equiv \gamma^{\mu\nu} k_\nu, \tag{3.32}$$

and substituting this back into (3.30) gives:

$$F(k_\mu) = \frac{1}{r(r-1)} \frac{\partial^2 F}{\partial k_\mu \partial k_\nu} k_\mu k_\nu = \frac{1}{r} \gamma^{\mu\nu} k_\mu k_\nu. \tag{3.33}$$

Note that when $r = 1$, one has $\gamma^{\mu\nu} = 0$.

If $r$ is a positive integer then one can continue this process until one reaches:

$$F(k_\mu) = \frac{1}{r!} \frac{\partial^r F}{\partial k_{\mu_1} \cdots \partial k_{\mu_r}} k_{\mu_1} \cdots k_{\mu_r}, \tag{3.34}$$

as one would expect from the theory of Taylor series.

Now, let us start with the fact that if $F[k] = 0$ and $F$ is homogeneous of degree $r$ in $k$ then, from Euler's formula, one must have:

$$k_\mu v^\mu = k_\mu \frac{\partial F}{\partial k_\mu} = rF[k] = 0. \tag{3.35}$$

If $k = \omega\, dt - k_i\, dx^i$ and $v^\mu = v^0 \partial_0 + v^i \partial_i$ then this gives:

$$k_i v^i = \omega v^0, \tag{3.36}$$

and dividing both sides by $\omega v^0$ gives:

$$s_i v_g^i = 1. \tag{3.37}$$

We can interpret this as meaning that if one regards $F[k]$ as a polynomial on $\mathbf{RP}^{3*}$, instead of $\mathbf{R}^{4*}$ then the characteristic hypersurface $F[k] = 0$ defines a surface $F[s] = 0$ in $\mathbf{RP}^{3*}$ that generalizes the Fresnel normal surface that pertains to the electromagnetic case. Similarly, the characteristic hypersurface $G[\mathbf{v}] = 0$ in $\mathbf{R}^4$ defines a surface $G[\mathbf{v}_g] = 0$ in $\mathbf{RP}^3$ that generalizes the Fresnel ray surface; these two surfaces are related by (3.37) as long as $F$ is homogeneous. This projective-geometric relationship between the Fresnel wave surface and the Fresnel ray surface was known in the quadratic case since the early part of the Twentieth Century (cf., Kommerell [**12**]), although eventually interest in the projective-geometric aspects of physical spacetime seemed to dwindle in favor of Riemannian geometry. Hopefully, the fact that projective geometry seems to be the natural pre-metric setting for wave motion will rekindle some of that early enthusiasm.

One must observe that when $F[k]$ is a homogeneous polynomial of degree $r$, the corresponding function $G[\mathbf{v}]$ will not generally be a polynomial, although it will be a homogeneous function of degree $r/(r-1)$, since the function $\mathbf{v}(k)$ will be a homogeneous polynomial of degree $r-1$, which means that its inverse $k(\mathbf{v})$, which one must *assume* exists, will be homogeneous of degree $1/(r-1)$. Of course, in the quadratic case ($r = 2$), matters are simplified, since $\mathbf{v}(k)$ is linear, as well as its inverse, which makes $G[\mathbf{v}]$ quadratic along with $F[k]$. Hence, a light cone in a tangent space corresponds to a light cone in a cotangent space.

## 4 Null geodesics for homogeneous characteristic polynomials

In this section, we shall be concerned with the form that equations (3.21) take when the function $F$ is homogeneous [4] of degree $r$ in the coordinates $k_\mu$; i.e., on the fibers of $T^*M$. As we shall see, they essentially generalize the geodesic equations that one obtains for the metric $\gamma^{\mu\nu}$ in the quadratic, Lorentzian case.

---

[4] Hopefully, one has noticed, by now, that there are considerable grounds for confusion in the inconsistent use of the words "characteristic" and "homogeneous."

In the case of a function $F$ on $T^*M$ that is defined to be homogeneous of degree $r$ on the fibers, one can use (3.32) and (3.33) to put equations (3.21) into the form:

$$\frac{dx^\mu}{d\tau} = \gamma^{\mu\nu} k_\nu, \qquad \frac{dk_\mu}{d\tau} = -\frac{1}{r}\frac{\partial \gamma^{\kappa\nu}}{\partial x^\mu} k_\kappa k_\nu. \tag{4.1}$$

We shall show that the bicharacteristic vector field that they define on $T^*M$ generalizes the geodesic flow that a metric would define.

In these equations, one must be careful to note that although we are using a standard component notation borrowed from Riemannian geometry, nonetheless, the component functions $\gamma^{\mu\nu}$ are defined on the total space of $T^*M$, not the base space $M$, as usual. This has the effect of making $\gamma^{\mu\nu} = \gamma^{\mu\nu}(x^\alpha, k_\beta)$, in general. Hence, the metric that one defines by $\gamma = \gamma^{\mu\nu} dk_\mu dk_\nu$ does not pertain to covectors in the fibers of $T^*M$, but covectors in a sub-bundle of $T^*(T^*M)$. In order to convert this $\gamma$ into a doubly covariant tensor field on $M$ one must "pull it down" by means of a specified covector field $k$ on $M$. That is, one must fix the coordinates $k_\mu$ in the component functions and replace the $dk_\mu$ with $(\partial k_\mu/\partial x^\nu)\, dx^\nu$. The resulting metric on $T(M)$ has the components:

$$\gamma^{\mu\nu}(x) = \frac{\partial k_\alpha}{\partial x^\mu}\frac{\partial k_\beta}{\partial x^\nu} \gamma^{\alpha\beta}(x, k). \tag{4.2}$$

This does, of course, imply that the resulting metric will depend upon the choice of $k$.

Homogeneity, combined with (4.1), implies that:

$$0 = k_\mu \frac{dx^\mu}{d\tau} = \gamma^{\mu\nu} k_\mu k_\nu = \gamma_{\mu\nu} \frac{dx^\mu}{d\tau} \frac{dx^\nu}{d\tau}, \tag{4.3}$$

under the assumption that the matrix $\gamma^{\mu\nu}$ is invertible with an inverse that is described by $\gamma_{\mu\nu}$. Hence, the covector $k_\mu$ and the vector $dx^\mu/d\tau$ are both *isotropic* for the metric $\gamma$ defined by $F$. This implies that $dx^\mu/d\tau$ – hence, the integral curve through it – lies in the hyperplane that is annihilated by $k_\mu$.

We should point out that what we are doing is essentially a generalization of what one does in the geometry of *Finsler spaces* [13, 14], which generalize Riemannian geometry by defining a positive definite function on $T(M)$ that is homogeneous of degree one, and thus generalizes a Euclidian norm on the tangent spaces. The reason that we are not trying to force the formalism into that template is the fact that we are not generally dealing with a positive definite function for a dispersion polynomial, so taking the $r^{\text{th}}$ root to obtain homogeneity of degree one might give imaginary values, not non-negative real ones, and differentiation of $F$ tends to produce unwanted zero denominators. However, one can use Finsler geometry as a heuristic guide for the formulation of one's definitions.

In particular, we further specialize $F$ by demanding that the matrices of components $\gamma^{\mu\nu}$ be invertible, so that we are indeed dealing with something that can possibly define a metric.

This allows us to differentiate the first equation in (4.1), and we get:

$$\frac{d^2 x^\mu}{d\tau^2} = \left( \frac{\partial \gamma^{\mu\nu}}{\partial x^\kappa} \frac{dx^\kappa}{d\tau} + \frac{\partial \gamma^{\mu\nu}}{\partial k_\kappa} \frac{dk_\kappa}{d\tau} \right) \gamma_{\kappa\nu} \frac{dx^\kappa}{d\tau} + \gamma^{\mu\nu} \frac{dk_\nu}{d\tau}. \tag{4.4}$$

Substituting from the second one gives:

$$\frac{d^2 x^\mu}{d\tau^2} = -\frac{1}{2} \gamma^{\mu\nu} \left( \frac{\partial \gamma_{\nu\lambda}}{\partial x^\kappa} + \frac{\partial \gamma_{\kappa\nu}}{\partial x^\lambda} - \frac{2}{r} \frac{\partial \gamma_{\kappa\lambda}}{\partial x^\nu} \right) \frac{dx^\kappa}{d\tau} \frac{dx^\lambda}{d\tau} + \gamma_{\lambda\nu} \frac{\partial \gamma^{\mu\nu}}{\partial k_\kappa} \frac{dk_\kappa}{d\tau} \frac{dx^\lambda}{d\tau}, \tag{4.5}$$

in which we have also switched upper and lower indices on $\gamma$ according to the usual rule.

This results in a system of second-order ordinary differential equations for $x^\mu$ of the form:

$$\frac{d^2 x^\mu}{d\tau^2} + \Gamma^\mu_{\kappa\lambda} \frac{dx^\kappa}{d\tau} \frac{dx^\lambda}{d\tau} = B^\mu_{\kappa\lambda} \frac{dx^\kappa}{d\tau} \frac{dx^\lambda}{d\tau} + C^{\mu\kappa}_\lambda \frac{dk_\kappa}{d\tau} \frac{dx^\lambda}{d\tau}, \tag{4.6}$$

in which we have introduced the notations:

$$\Gamma^\mu_{\kappa\lambda} = \tfrac{1}{2} \gamma^{\mu\nu} \left( \frac{\partial \gamma_{\nu\kappa}}{\partial x^\lambda} + \frac{\partial \gamma_{\nu\lambda}}{\partial x^\kappa} - \frac{\partial \gamma_{\kappa\lambda}}{\partial x^\nu} \right), \quad B^\mu_{\kappa\lambda} = \left( \frac{1}{r} - \frac{1}{2} \right) \gamma^{\mu\nu} \frac{\partial \gamma_{\kappa\lambda}}{\partial x^\nu}, \quad C^{\mu\kappa}_\lambda = \gamma_{\lambda\nu} \frac{\partial \gamma^{\mu\nu}}{\partial k_\kappa}. \tag{4.7}$$

The expressions $\Gamma^\mu_{\kappa\lambda}(x,k)$ take the form of the components of the Levi-Civita connection for the metric $\gamma$, except that, as before, they will be functions on a local coordinate chart on $T^*M$, not a local coordinate chart on $M$. The coefficients $B^\mu_{\kappa\lambda}$ clearly vanish when $F$ is homogeneous of degree 2, such as a metric tensor, and thus represent a correction to the Levi-Civita connection.

The components $C^{\mu\kappa}_\lambda$ also define a contribution to parallel translation that vanishes only for homogeneous $F$ of degree 2, except that they act on both the $dk_\mu / d\tau$, as well as the $dx^\mu / d\tau$. This contribution is referred to as the *Cartan connection* in Finsler geometry.

Since the contributions from the non-vanishing of the $B^\mu_{\kappa\lambda}$ and $C^{\mu\kappa}_\lambda$ also express the deviation of the proper acceleration of the curve from geodesic motion, we might regard them as one source of "quantum fluctuations about the classical extremals" that originates in the deviation of the spacetime electromagnetic dispersion law from a quadratic one, as might be a consequence of vacuum polarization. This would be in addition to the diffraction effects that are omitted by the geometrical optics approximation, which has long been recognized as being closely related to the classical limit of the quantum wave equations.

One should be careful not to conclude that we have succeeded in decoupling the $x^\mu$ coordinates from the $k_\mu$ coordinates in our system of differential equations. One cannot fix the $k_\mu$ arbitrarily in order to determine the $\Gamma^\mu_{\kappa\lambda}$ and $C^{\mu\kappa}_\lambda$ as functions of the $x^\mu$, since we had to make use of the equation:

$$k_\mu = \gamma_{\mu\nu}(x,k) \frac{dx^\nu}{d\tau} \tag{4.8}$$

in order to obtain (4.4). Hence, the equations for $k_\mu$ are still coupled to the equations for the $x^\mu$ by way of the $\gamma_{\mu\nu}(x, k)$, which depend upon both sets of independent variables.

The characteristic polynomials that are of most direct interest to physics are the ones that follow from the procedure described in section 2 when one starts with some – generally empirically defined – constitutive law for the medium. Hence, they are generally homogeneous polynomials in $k$ of degree 2, 4, and 6. We shall now specialize the above formulae for the quadratic and quartic cases.

### 4.1 Quadratic characteristic polynomials

Most commonly in the physics of waves one deals with a quadratic dispersion law of Lorentzian form:

$$g^{\mu\nu} k_\mu k_\nu = \begin{cases} 0 & \text{characteristic waves,} \\ \omega_0^2(x) & \text{non-characteristic waves,} \end{cases} \quad (4.9)$$

with:

$$g^{\mu\nu} k_\mu k_\nu = \omega^2 - \alpha^2 \kappa^2 = \omega^2 - \alpha^2(x) g^{ij}(x) k_i k_j . \quad (4.10)$$

The last term defines a Euclidian metric on $\Sigma^*$ by way of $g_s = g^{ij} \partial_i \otimes \partial_j$.

We have deliberately chosen not to include the factor $c^2$ to serve as a units conversion constant, as is customary, since the velocity of propagation of waves in a general wave medium is a property of the medium that must be derived from the dispersion law, and does not always lead to a unique constant, such as $c$, but only when the constitutive law of the medium is assumed to be linear (in the field strengths or infinitesimal strains), homogeneous, and isotropic. Hence, we have simply replaced that conversion factor with the coefficient $\alpha^2$, which is assumed to be a differentiable function on $M$.

As long as one is concerned with characteristic wave motion, for which $F[k]$ vanishes, one can represent a quadratic dispersion law as a degenerate form of a homogeneous dispersion law of degree $2n$ in which:

$$F[k] = (g(k, k))^n. \quad (4.11)$$

Since the zero locus of $F[k]$ is then identical with the zero locus of $g(k, k)$, one can simply replace a polynomial of the form (4.11) with a quadratic one of the form $F[k] = g(k, k)$.

If we return to equations (4.9) then we see that with $r = 2$, the metric components $\gamma_{\mu\nu}$ coincide with the $g_{\mu\nu}$ above – i.e., they are not functions of $k$ – the components $B^\mu_{\kappa\lambda}$ and $C^{\mu\kappa}_\lambda$ vanish, and the components $\Gamma^\mu_{\kappa\lambda}$ then represent the usual Levi-Civita connection that is associated with $g$. Hence, the geodesics that one obtains are the usual null geodesics of general relativity.

Now, let us compute the group velocity line field $\mathbf{v}_g$ that one obtains from the dispersion law and compare it to the phase velocity.

In the characteristic case, the dispersion law reduces to the linear form:

$$\omega = \alpha \kappa = \alpha \sqrt{g^{ij} k_i k_j} . \quad (4.12)$$

The spatial group velocity line field ($v_g^1, \ldots, v_g^n$) that is defined by such a dispersion law is obtained from:

$$v_g^i = \frac{\partial \omega}{\partial k_i} = \frac{\alpha}{\kappa} g^{ij} k_j \equiv \alpha n^i, \qquad (4.13)$$

in which we have introduced the spatial unit vector:

$$n^i = 1/\kappa \, k^i = 1/\kappa \, g^{ij} k_j. \qquad (4.14)$$

Hence, we can think of the scaling function $\alpha$ as being equal to:

$$\alpha(x) = v_g \equiv \sqrt{g_{ij} v_g^i v_g^j}, \qquad (4.15)$$

which makes it play the role of $c$ only in the direction of propagation that is defined by $n^i$.

From (4.13), one sees that the group velocity line field is collinear with the wave number line field ($k^1, \ldots, k^n$) that one obtains from the 1-form $k$ by means of the metric $g$.

Now, from (4.12), one also computes the phase velocity components to be:

$$v_p^i = \frac{\omega}{k_i} = \frac{\alpha \kappa}{k_i} = \frac{\alpha}{n_i}. \qquad (4.16)$$

As one sees, these components are not as geometrically meaningful as those of the slowness covector:

$$s_i = \frac{1}{\alpha} n_i = \frac{1}{\alpha \kappa} k_i. \qquad (4.17)$$

### 4.2 Factored quartic characteristic polynomials

When dealing with dispersion laws of even degree higher than two, one can make use of the fact that a common form that such homogeneous polynomials take in physics is the product of quadratic factors. As pointed out above, this factorization is universal to a broad class of nonlinear electrodynamical theories.

For instance, in the quartic case one might be dealing with a law of the form:

$$0 = F(k) = g(k,k)\bar{g}(k,k) = g^{\alpha\beta} \bar{g}^{\kappa\lambda} k_\alpha k_\beta k_\kappa k_\lambda. \qquad (4.18)$$

In this expression, we have suppressed the total symmetrization of the components as unnecessary, since the factored form is more convenient to work with. One generally assumes that the individual second-rank tensors $g$ and $\bar{g}$ are each of Lorentzian type.

Of particular interest is the case in which "extraordinary" Lorentzian metric $\bar{g}$ differs from the "ordinary" one $g$ only by a small perturbation:

$$\bar{g} = g + \varepsilon. \tag{4.19}$$

This situation might represent quantum fluctuations about a classical metric if the fluctuations were due to – say – vacuum birefringence resulting from vacuum polarization in the realm of high electromagnetic field strengths. For instance, the dispersion law that results from the Heisenberg-Euler Lagrangian (see [**4**, **15**]), which represents a one-loop effective Lagrangian for a photon field $F$ when one takes into account the formation of electron-positron pairs at high energy or high background field strengths, takes the form:

$$F(k) = (\eta(k, k) + \varepsilon_1 T(k, k))\,(\eta(k, k) + \varepsilon_2 T(k, k)), \tag{4.20}$$

in which $\eta$ represents the Lorentz scalar product on Minkowski space and $T$ represents the Faraday stress-energy-momentum tensor field that is associated with $F$.

For the situation that is described by (4.18), one would then have a dispersion polynomial of the form:

$$F(k) = g(k, k)^2 + g(k, k)\varepsilon(k, k). \tag{4.21}$$

One sees that the quadratic case treated previously can be regarded as obtained from the degenerate case in which $g = \bar{g}$ (i.e., $\varepsilon = 0$) by a redefinition of $F$ to $(g(k,k)\bar{g}(k,k))^{1/2}$.

The dispersion law for the characteristic wave case then decomposes into two distinct sub-cases: $g(k, k) = 0$, while $\bar{g}(k,k)$ is arbitrary, and the opposite case. Since the two metrics appear symmetrically, it is sufficient to treat either case as generic.

One sees that the geodesics of $F$ include the degenerate possibilities of null geodesics for $g$ and null geodesics for $\bar{g}$, which brings one back to the quadratic case. However, there is also the possibility of null geodesics that can move from one light cone to the other, as long as they intersect.

The characteristic equations for a quartic $F$ of the form (4.21) take the form:

$$\frac{dx^\mu}{d\tau} = \frac{1}{2}[\bar{g}(k,k)g^{\mu\nu} + g(k,k)\bar{g}^{\mu\nu}]k_\nu \equiv \gamma^{\mu\nu}(x, k)k_\nu, \tag{4.22a}$$

$$\frac{dk_\mu}{d\tau} = -\frac{1}{4}\left[\bar{g}(k,k)\frac{\partial g^{\kappa\lambda}}{\partial x^\mu} + g(k,k)\frac{\partial \bar{g}^{\kappa\lambda}}{\partial x^\mu}\right]k_\kappa k_\lambda = -\frac{1}{2}\frac{\partial \gamma^{\kappa\lambda}}{\partial x^\mu}k_\kappa k_\lambda, \tag{4.22b}$$

into which we have introduced the notation:

$$\gamma^{\mu\nu}(x, k) = \frac{1}{2}[\bar{k}^2 g^{\mu\nu} + k^2 \bar{g}^{\mu\nu}] \tag{4.23}$$

to connect with the formalism of section 4.1.

Note that although we are making the Finslerian assumption that the matrix $\gamma^{\mu\nu}$ is invertible, it should be clear that the inverse component matrix to $\gamma^{\mu\nu}$ does not have as elementary a form as the matrix being inverted.

In the perturbed degenerate quartic case, these equations take the form:

$$\frac{dx^\mu}{d\tau} = \{k^2 g^{\mu\nu} + \tfrac{1}{2}[\varepsilon(k,k)g^{\mu\nu} + k^2 \varepsilon^{\mu\nu}]\}k_\nu \equiv \gamma^{\mu\nu}k_\nu + \delta\gamma^{\mu\nu}k_\nu, \tag{4.24a}$$

$$\frac{dk_\mu}{d\tau} = -\frac{1}{2}k^2 \frac{\partial g^{\kappa\lambda}}{\partial x^\mu}k_\kappa k_\lambda - \frac{1}{4}\left[\varepsilon(k,k)\frac{\partial g^{\kappa\lambda}}{\partial x^\mu} + k^2 \frac{\partial \varepsilon^{\kappa\lambda}}{\partial x^\mu}\right]k_\kappa k_\lambda, \tag{4.24b}$$

which makes:

$$\gamma^{\mu\nu}(x,k) = k^2 g^{\mu\nu} + \tfrac{1}{2}[\varepsilon(k,k)g^{\mu\nu} + k^2 \varepsilon^{\mu\nu}] \tag{4.25}$$

in this case.

Something that becomes immediately clear is that when the perturbation $\varepsilon$ vanishes the right-hand sides of the characteristic equations vanish identically for null congruences, regardless of the nature of $g$. Consequently, one sees that it is more illuminating to use the quadratic dispersion law that is defined by $g$, rather than the quartic one that is defined by its square.

We naturally wish to contrast the form that the geodesic equations take in both cases above with the form that they take in the more familiar quadratic case. Since we have already developed the general formalism for homogeneous functions of $k$ above, we need only compute the components $B^\mu_{\kappa\lambda}$ and $C^{\mu\nu}_\kappa$ that we defined previously.

$$B^\mu_{\kappa\lambda} = -\frac{1}{8}\left[\bar{k}^2 g^{\mu\nu} + k^2 \bar{g}^{\mu\nu}\right]\frac{\partial \gamma_{\kappa\lambda}}{\partial x^\nu} \tag{4.26}$$

$$C^{\mu\nu}_\kappa = \gamma_{\kappa\lambda}[g^{\lambda\nu}\bar{g}^{\mu\alpha} + \bar{g}^{\lambda\nu}g^{\mu\alpha}]k_\alpha. \tag{4.27}$$

In the case (4.21) of a degenerate quartic characteristic polynomial that is perturbed by a quartic polynomial, the components take the form:

$$B^\mu_{\kappa\lambda} = \frac{1}{4}k^2 g^{\mu\nu}\frac{\partial \gamma_{\kappa\lambda}}{\partial x^\nu} - \frac{1}{8}\left[\varepsilon(k,k)g^{\mu\nu} + k^2 \varepsilon^{\mu\nu}\right]\frac{\partial \gamma_{\kappa\lambda}}{\partial x^\nu} \tag{4.29}$$

$$C^{\mu\nu}_\kappa = 2\gamma_{\kappa\lambda}g^{\lambda\nu}g^{\mu\alpha}k_\alpha + [\varepsilon^{\lambda\nu}g^{\mu\alpha} + g^{\lambda\nu}\varepsilon^{\mu\alpha}]k_\alpha. \tag{4.30}$$

In the case of the Heisenberg-Euler dispersion law, which takes the general form under consideration, with the unperturbed metric being the Lorentz metric $\eta^{\mu\nu}$ on Minkowski space, since the perturbation originates in the polarization of the electromagnetic vacuum in the realm of high field strengths, one sees that the right-hand side of (4.6) gives one a tangible origin to the concept of "quantum fluctuations about a classical extremal."   Of course, the actual computations are quite elaborate to carry out and difficult to interpret.

So far, we have only associated a kinematical quantity, namely, the velocity vector field **v**, for the motion of extended matter with a kinematical quantity, namely, the frequency-wave number 1-form $k$, for its motion as a wave.  We should also wish to associate dynamical quantities, as well.  However, since the usual process one encounters in wave mechanics seems to depend crucially upon the de Broglie relations, which equate the energy-momentum 1-form $p$ with $\hbar k$, an association that actually assumes that the wave function is describing the motion of a *point*, we shall defer further discussion until the sequel to this article.

## 5  Summary

In this part of the present two-part study, we have considered some of the issues that are associated with describing the motion of wavelike solutions of field equations that one defines by means of conservation laws for the fields and a constitutive law for the medium in which the fields exist.  We first showed that the same set of general field equations can describe mechanical waves and electromagnetic waves.

When one chooses a specific form for massless wavelike solutions of the field equations, the second-order field equation is replaced by a homogeneous polynomial equation $F[k] = 0$ for the physically acceptable wave covector fields $k$ that one calls the dispersion law for the medium and a set of first-order linear partial differential equations $k = d\phi$ for the phase function of the wave motion.  The level hypersurfaces of the phase function describe the motion of initial wave fronts.

By hypothesis, the function $F$ on the cotangent bundle $T^*M$ of the spacetime manifold $M$, when combined with its natural symplectic structure, defines a characteristic vector field on $T^*M$ whose flow generalizes the null geodesic flow that one obtains when $F$ is quadratic, as in the Lorentzian metric case.  The geodesic equations include terms in addition to the Levi-Civita terms that vanish when $F$ is quadratic and represent one source of "quantum fluctuations about classical extremals," in addition to the diffraction effects that are omitted by the geometrical approximation.  This is consistent with the fact that the dispersion law that is associated with the Heisenberg-Euler effective Lagrangian for electromagnetic waves, which includes the quantum contribution from vacuum polarization is actually of the quartic form, not the quadratic one that one obtains for linear and isotropic electromagnetic media.

Along with the change in the null geodesic equations, one also must contend with the fact that the diffeomorphism between cotangent spaces and corresponding tangent spaces that one obtains from $F$, and which we are calling "dimension-codimension duality," is not generally a linear isomorphism, as it is in the quadratic case, but a homogeneous polynomial map.  However, as long as $F[k]$ is a homogeneous function of $k$, one will always find that the relationship $k(\mathbf{v}) = k_\mu v^\mu = 0$ is valid when **v** is the velocity vector that is associated with the wave covector $k$ by this duality.

One finds that the usual Fresnel analysis comes about when one passes from the vector bundles $T^*M$ and $T(M)$ to their projectivizations $PT^*(M)$ and $PT(M)$; i.e., the sets of lines though the origins of their fibers.  The projection of $k$ is the slowness covector $s$, whose components are reciprocals of the phase velocity components, while the projection of **v** gives minus the usual group velocity vector.  As long as $F$ is homogeneous, one always has $s(\mathbf{v}_g) = 1$, which was known in the quadratic case.

In the sequel to this article, we shall address the manner in which the analysis of the foregoing presentation must be altered if one is to account for the motion of massive waves and their inhomogeneous dispersion laws.